\documentclass[aps, prd, 11pt, onecolumn, tightenlines, secnumarabic, superscriptaddress, nofootinbib]{revtex4-2}

\pdfoutput=1
\usepackage[utf8]{inputenc}
\usepackage[T1]{fontenc}
\usepackage{amsmath, amssymb, bm, slashed}
\usepackage{booktabs, tabularx}
\usepackage[labelformat=empty]{subcaption}
\usepackage{hyperref, cleveref}
\usepackage{graphicx}
\usepackage[normalem]{ulem}

\usepackage{feynmp-auto}

\begin{document}
	

\title{\texorpdfstring{Radiative Muon Mass Models and $(g-2)_\mu$}{Radiative Muon Mass Models and g-2}}
	
\author{Michael J.\ Baker}
\email{michael.baker@unimelb.edu.au}
\affiliation{ARC Centre of Excellence for Dark Matter Particle Physics, 
School of Physics, The University of Melbourne, Victoria 3010, Australia}
	
\author{Peter Cox}
\email{peter.cox@unimelb.edu.au}
\affiliation{ARC Centre of Excellence for Dark Matter Particle Physics, 
School of Physics, The University of Melbourne, Victoria 3010, Australia}
	
\author{Raymond R. Volkas}
\email{raymondv@unimelb.edu.au}
\affiliation{ARC Centre of Excellence for Dark Matter Particle Physics, 
School of Physics, The University of Melbourne, Victoria 3010, Australia}

	
\begin{abstract}
Recent measurements of the Higgs-muon coupling are directly probing muon mass generation for the first time.  We classify minimal models with a one-loop radiative mass mechanism and show that benchmark models are consistent with current experimental results.  We find that these models are best probed by measurements of $(g-2)_\mu$, even when taking into account the precision of Higgs measurements expected at future colliders.  The current $(g-2)_\mu$ anomaly, if confirmed, could therefore be a first hint that the muon mass has a radiative origin.
\end{abstract}

\maketitle

\section{Introduction}
\label{sec:introduction}

Radiative mass generation provides a compelling alternative to the tree-level mechanism of the Standard Model (SM), motivated by the fact that the fermion masses, with the exception of the top quark, are significantly smaller than the scale of electroweak symmetry breaking. This idea, that (some of) the SM fermion masses could be radiatively generated, has attracted significant attention over the years, and could form part of a broader understanding of the flavour sector. With the recent measurements of the Higgs-fermion couplings, and ongoing discussions weighing future Higgs factories, it is the ideal time to revisit these ideas and investigate the ability of current and future experiments to test radiative models.

The SM predicts a simple relation between the masses of the charged fermions and their couplings to the physical Higgs boson: $y_f = \sqrt{2} m_f/v$, where $v\approx246\,\text{GeV}$. This relation, which does not hold in radiative mass generation models, is being directly tested by Higgs measurements at the LHC. Until recently, these measurements were restricted to the third-family fermions; however, with the first evidence that the Higgs decays to muons~\cite{Sirunyan:2020two,Aad:2020xfq}, the mass generation mechanism of the second family is now being directly probed as well. For both families, the current data are consistent with the SM, but, as we shall see, also leave open the possibility of radiative mass generation. With increased precision at the (HL-)LHC and future Higgs factories, these measurements have the potential to distinguish between the two scenarios.

\begin{figure}[t]
  \begin{center}
    \includegraphics[width=0.8\textwidth]{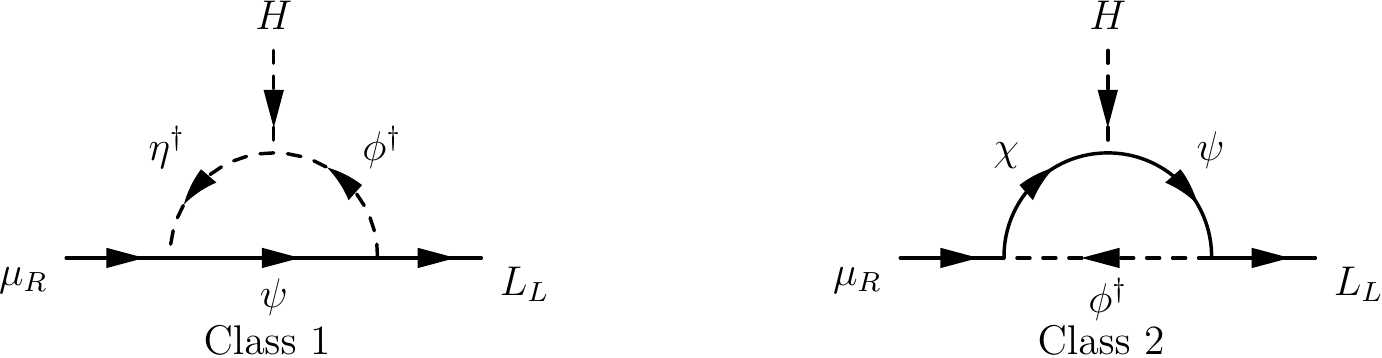}
  \end{center}
	\caption{One-loop diagrams that generate effective muon Yukawa couplings after integrating out the heavy, exotic fields.}
	\label{fig:feyn}
\end{figure}

In a recent work~\cite{Baker:2020vkh}, we performed a detailed analysis of one-loop radiative mass-generation for the $b$-quark and $\tau$-lepton. There we classified the minimal models containing new scalars and vector-like fermions, showing that they divide into two classes based on the field content and the diagram that generates the effective Yukawa coupling with the Higgs (see fig.\ref{fig:feyn}). In this paper, we apply this classification to explore one-loop radiative mass generation for the muon. 

Radiative models necessarily give new contributions to fermion anomalous magnetic moments~\cite{ hep-ph/9902443,hep-ph/0102122,0902.3360,1010.4485,1402.6415,1411.7362,1511.07458,2003.06633}. While this did not play a significant role in the phenomenology of the third-family models, it has important consequences in the case of the muon due to the high precision of the $(g-2)_\mu$ measurement. In fact, as we shall demonstrate, it provides the dominant constraint on these models. Furthermore, there is the interesting possibility that the $4.2\sigma$ discrepancy between the measurement~\cite{Abi:2021gix,Bennett:2006fi} and the SM prediction~\cite{Aoyama:2020ynm,1706.09436,1802.02995,1810.00007,1907.01556,1908.00921,1911.00367,1403.6400,hep-ph/0312226,1701.05829,1702.07347,1808.04823,1903.09471,1908.03331,1910.13432,1403.7512,1911.08123,1205.5370,10.3390/atoms7010028,hep-ph/0212229,1306.5546} could be explained by a radiative origin for the muon mass. The combination of increased data from the Fermilab experiment and improvements in lattice determinations of the HVP contribution (see~\cite{Borsanyi:2020mff}) should clarify the status of this anomaly in the future.  Regardless of whether the deviation persists, the anticipated increase in precision will have important implications for radiative mass generation models.

These minimal, one-loop models also generically possess a $U(1)_X$ symmetry that renders the lightest exotic state stable. If it is also neutral, this state may be a dark matter candidate, suggesting an intriguing link between the origins of fermion mass and dark matter.  Similar models with a viable dark matter candidate but a tree-level muon mass have been studied in~\cite{Kowalska:2017iqv,Calibbi:2018rzv,Kawamura:2020qxo,2006.07929,Jana:2020joi,Kowalska:2020zve}.

In the following sections we discuss the two classes of one-loop radiative models in turn. In each case, we begin by summarising the details of the model classification (which follows that of the $\tau$-lepton in Ref.~\cite{Baker:2020vkh}) before discussing the new contributions to the muon anomalous magnetic moment. We then focus on minimal benchmark models in order to perform a detailed analysis of the phenomenology and explore the sensitivity of future experiments.

\section{Class 1}
\label{sec:class-1}

\subsection{Class 1 -- Models}
\label{sec:1-models}

Class~1 models feature a single Dirac fermion $\psi$ and two scalars $\phi$ and $\eta$. An effective Yukawa coupling between the muon and the Higgs is then generated via the left diagram of \cref{fig:feyn}. The relevant Lagrangian terms for Class~1 models are
\begin{equation}
    \mathcal{L}_1 \supset\, - y_\phi \overline{L}_L \phi^\dagger \psi_R - y_\eta \overline{\psi}_L \eta \mu_R - a H \eta^\dagger \phi - m_\psi \overline{\psi}_L \psi_R + \mathrm{h.c.}\,,
        \label{eq:Lag1mu}
\end{equation}
where $L_L$ is the second-family left-handed lepton doublet, $\mu_R$ is the right-handed muon, and $H$ is the Higgs doublet. All couplings are taken to be real and positive without loss of generality. 

This class contains an infinite number of models, distinguished by the gauge quantum numbers of the new fields. The allowed Dynkin labels of the $SU(3)_C \times SU(2)_L$ representations and the hypercharges of the exotic fields are given in \cref{tab:class1}; the individual models are parameterised by the integers $a,b,c$ and the hypercharge $Y_\psi$. The above Lagrangian has two global $U(1)$ symmetries: generalised lepton number ($L$) and exotic particle number ($X$). The latter symmetry makes the lightest exotic stable, unless additional interactions are introduced between the exotics and the SM fields. There are also two softly broken $U(1)$ symmetries, $S_\psi$ and $S_a$, that forbid the tree-level Yukawa coupling between the muon and the Higgs.  They are broken by the fermion mass, $m_\psi$, and the trilinear coupling, $a$, respectively. 

For certain choices of the gauge quantum numbers, one of the exotic fields can be in a real representation of the SM gauge group. This field could then be a real scalar or Majorana fermion, with the appropriate modification of the above Lagrangian. This modifies the global symmetries: there remains a generalised lepton number and two softly broken $U(1)$ symmetries, but the exotic particle number symmetry is reduced to a $Z_2$. For generality, we shall always assume that the exotic fermion is Dirac and the scalars are complex. 

\begin{table}[t]
    \begin{tabular}{@{\hspace{1em}} c @{\hspace{2em}} c @{\hspace{2em}} c @{\hspace{2em}} c @{\hspace{2em}} c @{\hspace{2em}} c @{\hspace{2em}} c @{\hspace{2em}} c @{\hspace{1em}}}
        \toprule
        & $L_L$ & $\mu_R$ & $H$ & $\psi_L$ & $\psi_R$ & $\phi$ & $\eta$ \\
        \midrule
        $SU(3)_C$ & $(0,0)$ & $(0,0)$ & $(0,0)$ & $(a,b)$ & $(a,b)$ & $(a,b)$ & $(a,b)$ \\
        $SU(2)_L$ & $(1)$ & $(0)$ & $(1)$ & $(c)$ & $(c)$ & $(|c\pm1|)$ & $(c)$ \\
        $Y$ & $-\frac{1}{2}$ & $-1$ & $\frac{1}{2}$ & $Y_\psi$ & $Y_\psi$ & $Y_\psi\! +\! \frac{1}{2}$ & $Y_\psi\! +\! 1$ \\
        \midrule
        $L$ & $1$ & $1$ & $0$ & $0$ & $0$ & $-1$ & $-1$ \\
        $X$ & $0$ & $0$ & $0$ & $1$ & $1$ & $1$ & $1$ \\
        \midrule
        $S_\psi$ & $0$ & $1$ & $0$ & $1$ & $0$ & $0$ & $0$ \\
        $S_a$ & $0$ & $1$ & $0$ & $0$ & $0$ & $0$ & $-1$ \\
        \bottomrule
    \end{tabular}
    \caption{Quantum numbers of fields for Class~1 models. The first two lines give the allowed Dynkin labels for the non-Abelian groups.}
    \label{tab:class1}
\end{table}

\subsection{Class 1 -- Radiative Mass Generation and the Muon Anomalous Magnetic Moment}
\label{sec:1-mass-g-2}

Electroweak symmetry breaking leads to mixing of the exotic scalars via the $a$-term in \cref{eq:Lag1mu}. We denote the scalar mass eigenstates by $\tilde{\phi}$ and $\tilde{\eta}$, with the mixing angle given by
\begin{equation} \label{eq:mixing1}
    \sin(2\theta) = \frac{\sqrt{2} a v}{m_2^2 - m_1^2} \,,
\end{equation}
where $\theta \in [0,\tfrac{\pi}{2}]$. The eigenvalues are
\begin{align}
    m_1^2 \equiv m_{\tilde{\phi}}^2 &= \frac{1}{2}\left(m_\eta^2 + m_\phi^2 - \sqrt{(m_\eta^2 - m_\phi^2)^2+2 a^2 v^2}\right) \,, \\
    m_2^2 \equiv m_{\tilde{\eta}}^2 &= \frac{1}{2}\left(m_\eta^2 + m_\phi^2 + \sqrt{(m_\eta^2 - m_\phi^2)^2+2 a^2 v^2}\right) \,,
\end{align}
where $m_1 \leq m_2$. In general, there are $n_2\equiv\min(d_2(\phi),d_2(\eta))$ pairs of mixed states, with $d_2$ the dimension of the $SU(2)_L$ representation. The larger $SU(2)_L$ multiplet of $\phi$ and $\eta$ contains one state that does not mix. 

\begin{figure}[t]
  \begin{center}
    \includegraphics[width=0.8\textwidth]{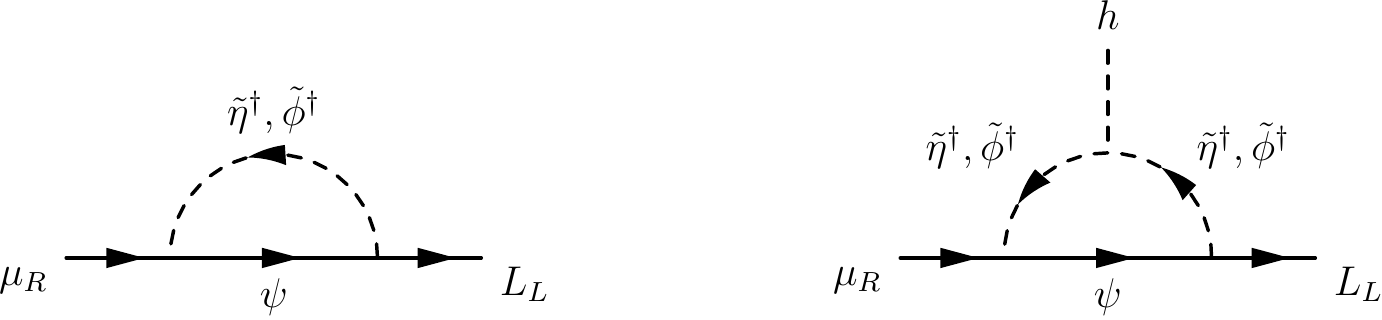}
  \end{center}
	\caption{One-loop diagrams that generate the muon mass and effective Yukawa coupling after electroweak symmetry breaking in the Class~1 models.}
	\label{fig:feyn-class1-after-ewsb}
\end{figure}

The one-loop muon mass, \cref{fig:feyn-class1-after-ewsb} (left), and the effective Yukawa coupling between the muon and the physical Higgs boson, \cref{fig:feyn-class1-after-ewsb} (right), are given by
\begin{align} \label{eq:1-mmu}
    m_\mu &= \frac{y_\phi y_\eta}{16\pi^2} 
    \frac{v}{\sqrt{2}}
    \frac{a m_\psi}{m_1 m_2} n_2 d_3 F\left(x_1, x_2\right) \,,\\
    \label{eq:1-eff-yuk}
    y_\mu^\text{eff}(p_h^2 = 0) &= \frac{\sqrt{2}m_\mu}{v} \left[ \cos^2(2\theta)  + \frac{1}{2} \sin^2(2\theta) \frac{\sqrt{x_1 x_2}}{F(x_1,x_2)} \left( \frac{F(x_1)}{x_1} + \frac{F(x_2)}{x_2} \right) \right] \,,
\end{align}
where $x_{1,2} = m_{1,2}^2/m_\psi^2$, and $d_3$ is the dimension of the $SU(3)_C$ representation of the exotics. The loop function, $F$, is given in~\cite{Baker:2020vkh}.  While for brevity we have written the effective Yukawa coupling at $p_h^2=0$, we use the $p_h^2=m_h^2$ expression in our numerical work.  We have also used \cref{eq:1-mmu} to write it in terms of the physical muon mass and highlight the violation of the SM relation $m_\mu = y_\mu v/\sqrt{2}$. Notice that \cref{eq:1-eff-yuk} is independent of $n_2$ and $d_3$, so the predicted deviation from the SM is the same for all Class~1 models.

The leading contribution to the anomalous magnetic moment of the muon, generated by attaching photons to the electrically charged internal legs of \cref{fig:feyn-class1-after-ewsb} (left), is
\begin{multline} \label{eq:1-g-2}
    \delta a_\mu = \frac{m_\mu^2}{m_\psi^2} \left(\frac{x_1\ln x_1}{1-x_1} - \frac{x_2\ln x_2}{1-x_2}\right)^{-1}  \bigg[ \frac{3x_1 - 1}{(1 - x_1)^2} - \frac{3x_2 - 1}{(1 - x_2)^2} + \frac{2x_1^2 \ln{x_1}}{(1 - x_1)^3} - \frac{2x_2^2 \ln{x_2}}{(1 - x_2)^3} \\
    + 2\overline{Q}_S \left(  \frac{1}{1 - x_1} - \frac{1}{1 - x_2} + \frac{x_1 \ln x_1}{(1 - x_1)^2} - \frac{x_2 \ln x_2}{(1 - x_2)^2} \right) \bigg] + \cdots \,,
\end{multline}
where $\overline{Q}_S=\frac{1}{n_2} \sum_{i=1}^{n_2} Q_i$ is the average of the charges of the mixed scalar states\footnote{We have neglected the small mass-splitting of scalar states with different charges due to electroweak corrections.}. In deriving this expression we have used \cref{eq:1-mmu}, with the result that the leading contribution to $a_\mu \equiv \frac{1}{2}(g-2)_\mu$ is independent of the couplings $y_{\phi}$, $y_\eta$, and the $SU(3)$ representation of the exotics. The `$\ldots$' denotes sub-leading terms arising from a chirality-flip on the muon line in \cref{fig:feyn-class1-after-ewsb} (left). These are formally of higher-loop order since the muon mass is radiatively generated. In our numerical results we include these terms, and also the log-enhanced two-loop QED corrections from~\cite{Degrassi:1998es}, with the new physics scale taken as the geometric mean of $m_1$, $m_2$ and $m_\psi$.

\subsection{Class~1 -- Phenomenology}
\label{sec:1-pheno}

To analyse the phenomenology of these models we consider those with the smallest gauge representations ($a=b=c=0$ in \cref{tab:class1}) for two hypercharge choices.  The relevant group theory factors in \cref{eq:1-mmu} are then $n_2 = d_3 = 1$.  The first Class~1 model we study is\footnote{This model was briefly discussed in~\cite{1402.6415}, although we find a different expression for the effective Yukawa coupling.}
\begin{equation}
    \psi_{L,R} \sim (\mathbf{1} ,\mathbf{1}, 0)\,, \quad \phi \sim (\mathbf{1}, \mathbf{2}, \tfrac{1}{2}) \,, \quad \eta \sim (\mathbf{1}, \mathbf{1}, 1) \,,
\end{equation}
where we have taken $Y_\psi=0$.  If $\psi$ is the lightest exotic, it may be a dark matter candidate.  While $\phi$ also contains a neutral state, it will typically be heavier than the lightest charged $\phi-\eta$ mixed state.  There is no dark matter candidate if there are additional interactions between the exotics and SM fields that completely break the $U(1)_X$ symmetry, for example $\overline{\psi}_R H L_L$, $\phi H^\dagger$,  $\eta \overline{L_L^c} L_L$.  If such terms are present there may be further constraints in addition to those discussed below.

The second Class 1 model we study is
\begin{equation}
    \psi_{L,R} \sim (\mathbf{1} ,\mathbf{1}, -1)\,, \quad \phi \sim (\mathbf{1}, \mathbf{2}, -\tfrac{1}{2}) \,, \quad \eta \sim (\mathbf{1}, \mathbf{1}, 0) \,,
\end{equation}
where we have taken $Y_\psi=-1$.  In this model the lighter mixed scalar, $\tilde{\phi}$, may be a dark matter candidate.  Alternatively, the $U(1)_X$ symmetry may be completely broken by additional terms such as $\overline{\psi}_R H^\dagger L_L$ or $\phi H$.

\begin{figure}[ht]
    \centering
    \includegraphics[height=0.45\textwidth]{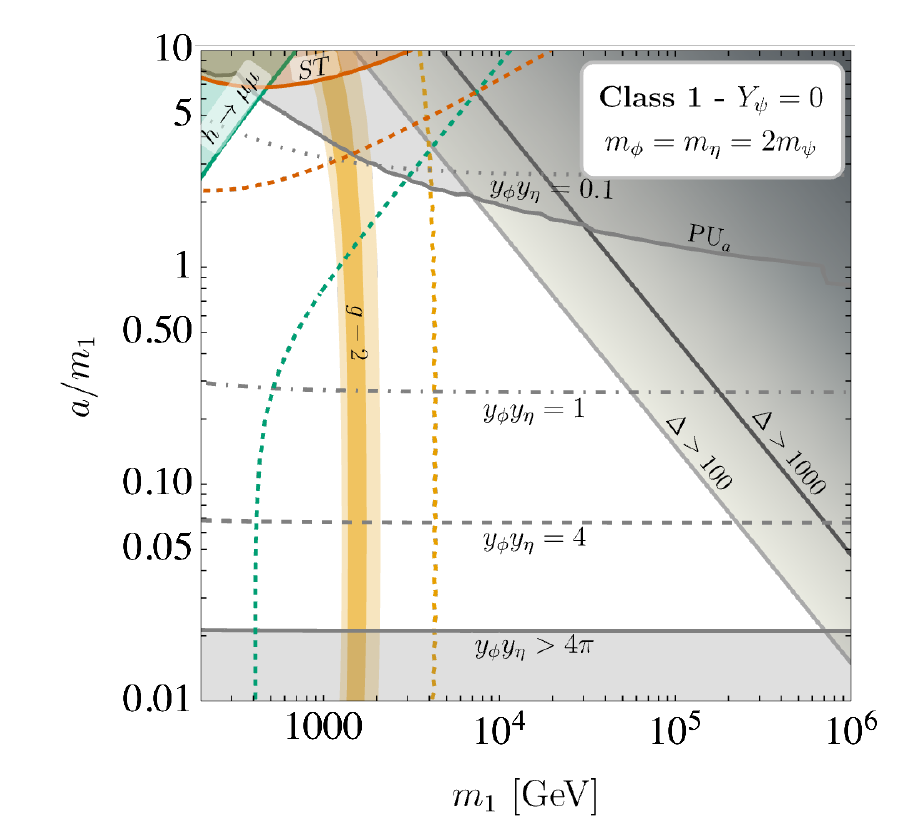}
    \includegraphics[height=0.45\textwidth]{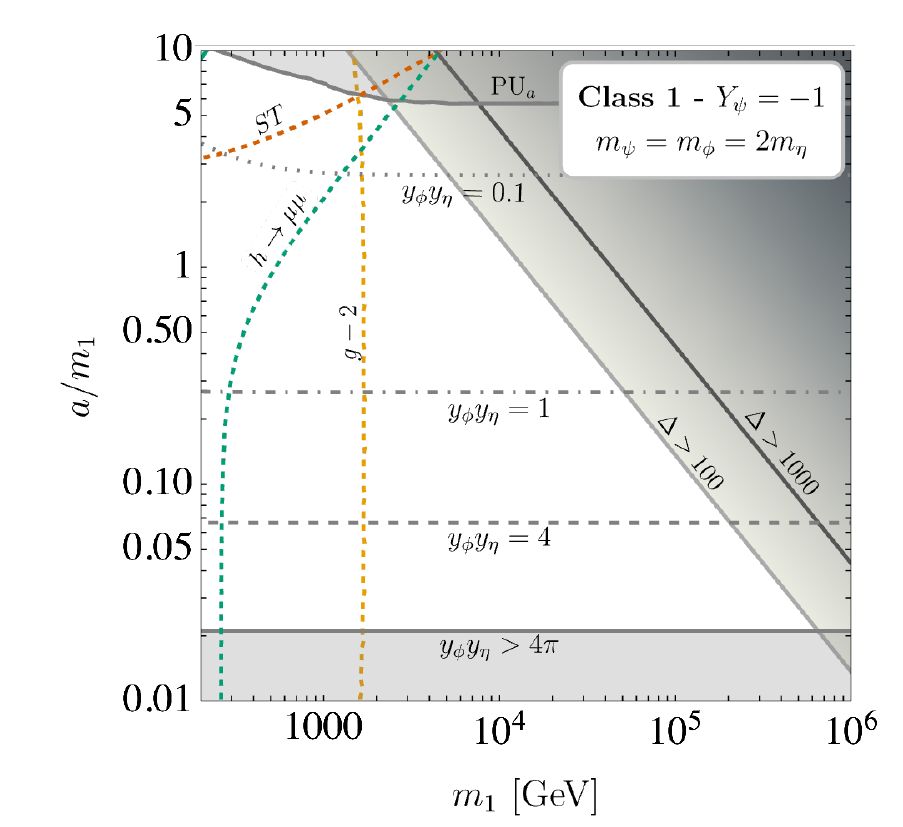}
	\caption{Theoretical (grey) and experimental (coloured) constraints on the Class~1 benchmark models.  The product of Yukawas required to reproduce the observed muon mass is shown by contours at $y_\phi y_\eta \in \{0.1, 1, 4, 4\pi\}$.  The region where perturbative unitarity of the trilinear coupling is violated is indicated by $\text{PU}_a$.  The degree of fine-tuning is shown by contours at $\Delta \in \{100,1000\}$ using the measure defined in~\cite{Baker:2020vkh}.  The orange bands show the region that fits the $(g-2)_\mu$ anomaly at 1 and 2~$\sigma$, while the green and red regions are experimentally excluded at $2\sigma$ by the Higgs-muon signal strength and the $S$ and $T$ parameters, respectively.  Coloured dotted contours show the corresponding $2\sigma$ projected future sensitivities.}
	\label{fig:1mu}
\end{figure}

There are several theoretical and experimental constraints which necessarily result from the fields and couplings required to generate the muon mass.  In \cref{fig:1mu} we show these constraints for the two models, where we fix the product\footnote{Of all the observables, only the sub-leading contributions to $(g-2)_\mu$ depend on $y_\phi$ and $y_\eta$ independently. We fix $y_\phi=y_\eta$, but this assumption has a negligible impact on the figure.} $y_\phi y_\eta$ by requiring that the observed muon mass be generated radiatively.  In the left panel, showing the $Y_\psi=0$ model, we take a parameter slice with maximal mixing ($m_\eta = m_\phi$) and where $\psi$ is the lightest exotic.  In the right panel, for the $Y_\psi = -1$ model, we show a slice with smaller mixing and the lightest exotic is $\eta$-like.

Perturbative unitarity places limits on how large the Yukawa couplings, $y_\phi$ and $y_\eta$, and the trilinear coupling $a$ can be.  Combining $y_\phi, y_\eta  \lesssim \sqrt{4\pi}$ with the muon mass requirement leads to the bound $a \gtrsim 0.02\,m_1$.  The perturbative unitarity bound on the trilinear is calculated using the \texttt{SARAH}/\texttt{SPheno} framework~\cite{Porod:2003um,Porod:2011nf,Staub:2008uz,Staub:2009bi,Staub:2010jh,Staub:2012pb,Staub:2013tta} using the default partial diagonalisation treatment of the $t$- and $u$-channel poles~\cite{Goodsell:2018tti,Goodsell:2020rfu}.   The difference in this bound between the two plots is driven by the parameter slices, rather than the differing hypercharges.  When the new physics scales become large there is a fine-tuning problem for the Higgs mass, as discussed in~\cite{Baker:2020vkh}, which disfavours this region of parameter space.  Within the theoretically preferred region of parameter space we see that the observed muon mass can be obtained for a product of Yukawas $y_\phi y_\eta \lesssim 0.1$ for small $m_1$; alternatively, $m_1$ can be as heavy as $\mathcal{O}(100\,\text{TeV})$ for larger Yukawas.

In the left-hand plot of \cref{fig:1mu} we see that the muon $g-2$ anomaly, $\Delta a_\mu = (251 \pm 59) \times 10^{-11}$~\cite{Abi:2021gix}, can be explained for $m_1 \approx 1-2\,\text{TeV}$, almost independently of the trilinear coupling $a$.  This independence arises because the leading contribution to the anomalous magnetic moment depends on $a$ only via $m_{1,2}$.  In the future, the Fermilab experiment is ultimately expected to achieve a precision of $16 \times 10^{-11}$~\cite{Chapelain:2017syu}. If the anomaly disappears, this measurement (shown by the dashed orange line) will be able to probe the parameter space up to $m_1 \approx 4\,\text{TeV}$.  The $Y_\psi=-1$ model (\cref{fig:1mu} right) can not fit the anomaly, since \cref{eq:1-g-2} can only contribute with the correct sign if $\overline{Q}_S>0$.  The only exception is if there exist regions of parameter space where the sub-leading terms in fact dominate and are large enough to explain the deviation.  Hence, if the anomaly is confirmed it would strongly disfavour this model. On the other hand, if the anomaly disappears, future measurements will be able to probe the parameter space up to $m_1 \approx 2\,\text{TeV}$.

A naive combination of the CMS~\cite{Sirunyan:2020two} and ATLAS~\cite{Aad:2020xfq} results for the Higgs-muon signal strength yields $\mu = 1.2\pm0.3$. \Cref{fig:1mu} shows that this measurement currently only constrains the parameter space of the $Y_\psi = 0$ model, and only for small masses and large trilinear coupling.  While it rules out theoretically allowed parameter space, it does not probe the region favoured by the $(g-2)_\mu$ anomaly.  Ultimately, a future collider such as the FCC may be able to measure the $h\bar{\mu}\mu$ coupling with a precision of $\sim0.4\%$~\cite{deBlas:2019rxi}. If the $(g-2)_\mu$ anomaly persists, the FCC would be able to probe the motivated region down to $a\approx m_1 \approx 1-2\,\text{TeV}$.  If the anomaly disappears, then future measurements of the muon $g-2$ will provide significantly better sensitivity than FCC measurements of the Higgs-muon coupling.  In the parameter slice we have taken for the $Y_\psi = -1$ model, the $h\to\bar{\mu}\mu$ and $(g-2)_\mu$ sensitivities are somewhat weaker, due to the larger fermion mass.  

The $Y_\psi = 0$ model also predicts a deviation in the Higgs-photon coupling, but both the current constraint~\cite{ATLAS:2020pvn,CMS:2020omd} and the future sensitivity are significantly weaker than for $h\bar{\mu}\mu$, so we do not plot them.  There is no such effect in the $Y_\psi = -1$ model since the mixed scalars, which couple to the Higgs, are electrically neutral.

These models also give corrections to the electroweak $S$ and $T$ parameters, see Ref.~\cite{Baker:2020vkh}. \Cref{fig:1mu} shows that these are, and will be, relatively weak probes of the parameter space. We use the electroweak fit and the future FCC-ee projections, both with $U=0$, from Ref.~\cite{1608.01509}.  While these models also lead to deviations in the $Z\bar{\mu}\mu$ coupling, the current limits~\cite{hep-ex/0509008} and FCC-ee projections~\cite{deBlas:2019wgy} are even less constraining than the Higgs observables, so we do not plot them.

There are also bounds from direct searches for the exotic states. In the $Y_\psi=0$ model these come from smuon searches~\cite{1908.08215}. If the scalar mixing is negligible, the limits on left- (right-) handed smuons can be directly applied to $\phi$ ($\eta$). This leads to the bounds $m_\phi \gtrsim 560\,\text{GeV}$ (provided $m_\psi \lesssim 300\,\text{GeV}$) and $m_\eta \gtrsim 450\,\text{GeV}$ (for $m_\psi \lesssim 150\,\text{GeV}$). This assumes that the $U(1)_X$ symmetry is (approximately) preserved and $\psi$ is collider-stable. For the $Y_\psi=-1$ model, recasting this search should lead to a similar bound on $m_\psi$, assuming $m_\psi > m_1$. At a future 100\,TeV proton collider, similar searches are projected to reach masses around 1.2\,TeV~\cite{Baker:2018uox}, assuming an integrated luminosity of 20\,ab$^{-1}$. Comparable mass reach could be obtained at a multi-TeV $e^+e^-$ collider such as CLIC~\cite{1812.02093}, which is expected to have sensitivity to masses close to half the centre-of-mass energy. Muon colliders could provide a means to ultimately reach even higher mass scales~\cite{2006.16277,Buttazzo:2020eyl,2012.03928,Capdevilla:2021rwo}.

In the absence of any additional terms that completely break the $U(1)_X$ exotic stabilising symmetry, both of these models contain dark matter candidates.  In the $Y_\psi = 0$ model, $\psi$ may constitute dark matter if it is the lightest exotic particle.  Its relic abundance would be set by annihilation into muons via $t$-channel $\phi$ and $\eta$ exchange, and possibly by co-annihilation effects if the $\psi$ mass is close to a scalar mass.  If $\psi$ is Dirac then there are expected to be significant constraints from direct detection experiments, arising at loop-level~\cite{1109.3516,1401.6457,1402.6696,1503.03382,Baker:2018uox}.  However, these limits can be evaded if $\psi$ is Majorana as the scattering rate becomes velocity suppressed.  This model, with Majorana $\psi$ but a tree-level muon mass, has been found to fit the $(g-2)_\mu$ anomaly and provide a viable dark matter candidate in~\cite{Kawamura:2020qxo}.  It would be very interesting to investigate whether a radiative muon mass, the $(g-2)_\mu$ anomaly, and viable dark matter can all be accommodated simultaneously.
 
In the $Y_\psi = -1$ model the lighter mixed scalar is a dark matter candidate.  The relic abundance would be set by $t$-channel $\psi$ exchange, electroweak processes, and possibly co-annihilation.  Unless the mixing angle is small, $Z$ exchange will lead to stringent direct detection constraints. These could be weakened if $\eta$ were taken to be real or if there was a mass splitting between the CP-even and CP-odd components~\cite{Calibbi:2018rzv} (note, however, that this would also impact the other phenomenology discussed above).

Finally, let us briefly comment on the phenomenology of non-minimal models in this class, where the exotic fields are in larger representations of the non-Abelian groups.  Interestingly, the effective Higgs-muon Yukawa coupling is model-independent (see \cref{eq:1-eff-yuk}); however, in models with coloured scalars there are new contributions to the Higgs-gluon coupling, which modify the production cross-section and hence the Higgs-muon signal strength.  In addition, the contribution to the Higgs-photon coupling will generally be larger in non-minimal models.  The size of the contribution to $(g-2)_\mu$ is expected to be comparable in all models (provided the hypercharges are $\mathcal{O}(1)$); this can be seen from the fact that the only model-dependence in \cref{eq:1-g-2} is via the average of the mixed scalar charges, $\overline{Q}_S$. The leading contribution has the correct sign to explain the anomaly when $\overline{Q}_S \geq 1$ and may do if $0<\overline{Q}_S<1$, depending on the masses of the exotics.  Models with coloured exotics would be subject to significantly stronger bounds from direct searches, and give larger contributions to the $S$ and $T$ parameters.

\section{Class 2}
\label{sec:class-2}

\subsection{Class 2 -- Models}
\label{sec:2-models}

In Class~2 models the effective Yukawa coupling between the muon and the Higgs is generated by the right diagram in \cref{fig:feyn}. These models contain two new fermions, $\psi$ and $\chi$, and a new scalar, $\phi$. The Lagrangian contains the following terms,
\begin{equation}
     \mathcal{L}_2 \supset\, -y_\psi \overline{L}_L \phi^\dagger \psi_R - y_\chi \overline{\chi}_L \phi \mu_R - y_H \overline{\psi}_L H \chi_R - m_\psi \overline{\psi}_L \psi_R - m_\chi \overline{\chi}_L \chi_R + \mathrm{h.c.}\,,
     \label{eq:2-lag}
\end{equation}
where we can take all couplings to be real and positive without loss of generality. 

The allowed gauge representations of the exotic fields in Class~2 models are given in \cref{tab:class2}. As in Class~1, there are two exact global symmetries, $L$ and $X$. The symmetries $S_\psi$ and $S_\chi$ forbid the Higgs-muon Yukawa coupling and are softly broken by $m_\psi$ and $m_\chi$ respectively. In specific models one of the exotic fields may again be in a real representation; for generality, we always assume that the fermions are Dirac and that $\phi$ is complex.

\begin{table}[ht]
    \begin{tabular}{@{\hspace{1em}} c @{\hspace{2em}} c @{\hspace{2em}} c @{\hspace{2em}} c @{\hspace{2em}} c @{\hspace{2em}} c @{\hspace{2em}} c @{\hspace{2em}} c @{\hspace{2em}} c @{\hspace{1em}}}
        \toprule
        & $L_L$ & $\mu_R$ & $H$ & $\psi_L$ & $\psi_R$ & $\chi_L$ & $\chi_R$ & $\phi$ \\
        \midrule
        $SU(3)_C$ & $(0,0)$ & $(0,0)$ & $(0,0)$ & $(a,b)$ & $(a,b)$ & $(a,b)$ & $(a,b)$ & $(a,b)$ \\
        $SU(2)_L$ & $(1)$ & $(0)$ & $(1)$ & $(|c\pm1|)$ & $(|c\pm1|)$ & $(c)$ & $(c)$ & $(c)$ \\
        $Y$ & $-\frac{1}{2}$ & $-1$ & $\frac{1}{2}$ & $Y_\psi$ & $Y_\psi$ & $Y_\psi\! -\! \frac{1}{2}$ & $Y_\psi\! -\! \frac{1}{2}$ & $Y_\psi\! +\! \frac{1}{2}$ \\
        \midrule
        $L$ & $1$ & $1$ & $0$ & $0$ & $0$ & $0$ & $0$ & $-1$ \\
        $X$ & $0$ & $0$ & $0$ & $1$ & $1$ & $1$ & $1$ & $1$ \\
        \midrule
        $S_\chi$ & $0$ & $1$ & $0$ & $0$ & $0$ & $1$ & $0$ & $0$ \\
        $S_\psi$ & $0$ & $1$ & $0$ & $0$ & $-1$ & $0$ & $0$ & $-1$ \\
        \bottomrule
    \end{tabular}
    \caption{Quantum numbers of fields for Class~2 models.  The first two lines give the allowed Dynkin labels for the non-Abelian groups.}
    \label{tab:class2}
\end{table}

\subsection{Class 2 -- Radiative Mass Generation and the Muon Anomalous Magnetic Moment}
\label{sec:2-mass-g-2}

After electroweak symmetry breaking the $y_H$ term in \cref{eq:2-lag} mixes the fermions $\psi$ and $\chi$ to give mass eigenstates denoted by $\tilde{\psi}$ and $\tilde{\chi}$. The mass eigenvalues are
\begin{align}
    m_1^2 \equiv m_{\tilde{\psi}}^2 &= \frac{1}{4} \left(2m_\psi^2 + 2m_\chi^2 + y_H^2 v^2 - \sqrt{4(m_\chi^2 - m_\psi^2)^2 + y_H^2 v^2 (4m_\psi^2 + 4m_\chi^2 + y_H^2 v^2)} \right) \,, \\
    m_2^2 \equiv m_{\tilde{\chi}}^2 &= \frac{1}{4} \left( 2m_\psi^2 + 2m_\chi^2 + y_H^2 v^2 + \sqrt{4(m_\chi^2 - m_\psi^2)^2 + y_H^2 v^2 (4m_\psi^2 + 4m_\chi^2 + y_H^2 v^2)} \right) \,.
\end{align}
The mixing angles for the left- and right-handed fields, with $\theta_L,\theta_R \in [0,\tfrac{\pi}{2}]$, are
\begin{align}
    \tan(2\theta_L) &= 2\sqrt{2}\, y_H v \frac{m_\chi}{2(m_\chi^2 - m_\psi^2) - y_H^2 v^2} \,, \\
    \tan(2\theta_R) &= 2\sqrt{2}\, y_H v \frac{m_\psi}{2(m_\chi^2 - m_\psi^2) + y_H^2 v^2} \,.
\end{align}
There are $n_2\equiv\min(d_2(\psi),d_2(\chi))$ pairs of mixed states, and the larger $SU(2)_L$ multiplet of $\psi$ and $\chi$ contains one state that does not mix.

\begin{figure}[t]
  \begin{center}
    \includegraphics[width=0.8\textwidth]{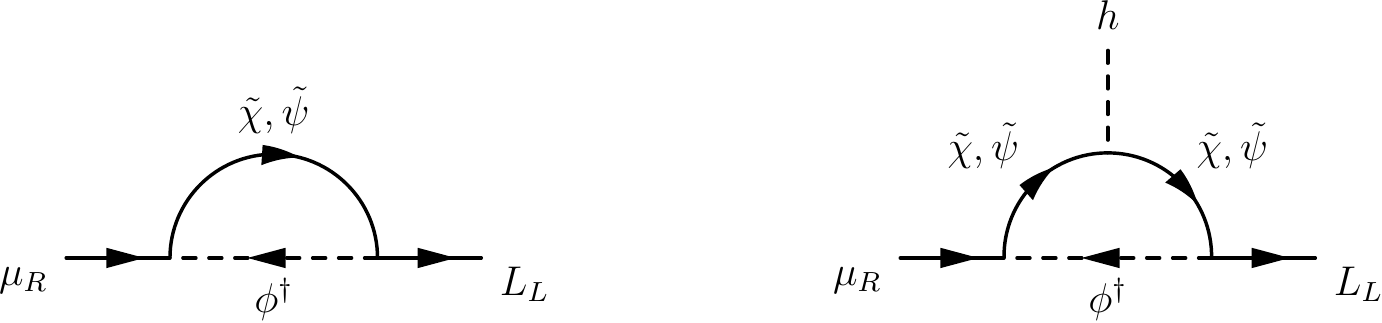}
  \end{center}
	\caption{One-loop diagrams that generate the muon mass and effective Yukawa coupling after electroweak symmetry breaking in the Class~2 models.}
	\label{fig:2-feyn-after-ewsb}
\end{figure}

The one-loop radiative muon mass and effective Yukawa coupling, generated by the Feynman diagrams in \cref{fig:2-feyn-after-ewsb}, are given by
\begin{align} \label{eq:2-mmu}
    m_\mu &= \frac{y_\psi y_\chi}{16\pi^2} \frac{y_H v}{\sqrt{2}} n_2 d_3 F\left(x_1,x_2\right) \,,\\
    \label{eq:2-eff-yuk}
    y_{\mu}^\text{eff}(p_h^2 = 0) &= \frac{\sqrt{2}m_\mu}{v} \Bigg[ 1 + \sin\theta_L \cos\theta_L \sin\theta_R \cos\theta_R \bigg( \frac{G(x_1, x_2)}{F(x_1,x_2)} - \frac{x_1 + x_2}{\sqrt{x_1 x_2}} \bigg) \Bigg] \,,
\end{align}
where now $x_{1,2} = m_{1,2}^2/m_\phi^2$, $d_3$ is the dimension of the $SU(3)_C$ representation, and $F(x_1,x_2)$, $G(x_1,x_2)$ are defined in \cite{Baker:2020vkh}.  We have written the effective Yukawa in terms of the physical muon mass, to show how the SM relation $m_f = y_fv/\sqrt{2}$ is violated.  It is then clear that the effective Yukawa coupling is independent of the representations of the exotics.  While here we give the simpler $p_h^2 = 0$ expression for $y_{\mu}^\text{eff}$, in our numerical work we take $p_h^2 = m_h^2$.

The leading contribution to the anomalous magnetic moment of the muon, given by the Feynman diagram in \cref{fig:2-feyn-after-ewsb} (left) with photons attached to the electrically charged exotic particles, is
\begin{multline} \label{eq:2-g-2}
    \delta a_\mu = \frac{m_\mu^2}{m_\phi^2} \left(\frac{x_1\ln x_1}{1-x_1} - \frac{x_2\ln x_2}{1-x_2}\right)^{-1}  \bigg[ \frac{1 + x_1}{(1 - x_1)^2} - \frac{1 +  x_2}{(1 - x_2)^2} + \frac{2x_1 \ln x_1}{(1 - x_1)^3} - \frac{2x_2 \ln x_2}{(1 - x_2)^3} \\
    - 2\overline{Q}_F \left( \frac{1}{1 - x_1} - \frac{1}{1 - x_2 } + \frac{\ln x_1}{(1 - x_1)^2} - \frac{\ln x_2}{(1 - x_2)^2} \right) \bigg] + \cdots \,,
\end{multline}
where $\overline{Q}_F=\frac{1}{n_2} \sum_{i=1}^{n_2} Q_i$ is the average of the charges of the mixed fermion states.   After using \cref{eq:2-mmu}, the leading contribution to $(g-2)_\mu$ does not depend on the couplings $y_\psi$ and $y_\chi$ or on $d_3$.  The sub-leading terms denoted by `$\ldots$' again arise from a chirality-flip on the muon line and are formally of higher-loop order. In our numerical analysis we include these terms as well as the two-loop QED corrections~\cite{Degrassi:1998es}.

\subsection{Class~2 -- Phenomenology}
\label{sec:2-pheno}

The phenomenology of the Class~2 models is in many ways analogous to that of Class~1.  The models necessarily impact the same observables, and many of the features seen above will again appear here.

As benchmark models we again take the smallest gauge representations for the exotic fields ($a=b=c=0$ in \cref{tab:class2}) for two hypercharge choices.  For these models the group theory factors in \cref{eq:2-mmu} are $n_2 = d_3 = 1$.  The first model we study has
\begin{equation}
    \psi_{L,R} \sim (\mathbf{1},\mathbf{2},\tfrac{1}{2})\,, \quad \chi_{L,R} \sim (\mathbf{1},\mathbf{1},0) \,, \quad \phi \sim (\mathbf{1},\mathbf{1},1) \,,
\end{equation}
with $Y_\psi=\tfrac{1}{2}$.  The lightest mixed $\psi-\chi$ state may be a dark matter candidate. Alternatively, the $U(1)_X$ symmetry may be broken by additional terms such as $\overline{\chi}_R H L_L$ or $\phi \overline{L}_L^c L_L$.

The second model we consider is
\begin{equation}
    \psi_{L,R} \sim (\mathbf{1},\mathbf{2},-\tfrac{1}{2})\,, \quad \chi_{L,R} \sim (\mathbf{1},\mathbf{1},-1) \,, \quad \phi \sim (\mathbf{1},\mathbf{1},0) \,,
\end{equation}
corresponding to $Y_\psi=-\tfrac{1}{2}$.  In this case the scalar $\phi$ may be a dark matter candidate. Alternatively, the $U(1)_X$ symmetry may be broken by terms like $\overline{\psi}_L H e_R$ or $\phi |H|^2$.

\begin{figure}[ht]
    \centering
    \includegraphics[height=0.45\textwidth]{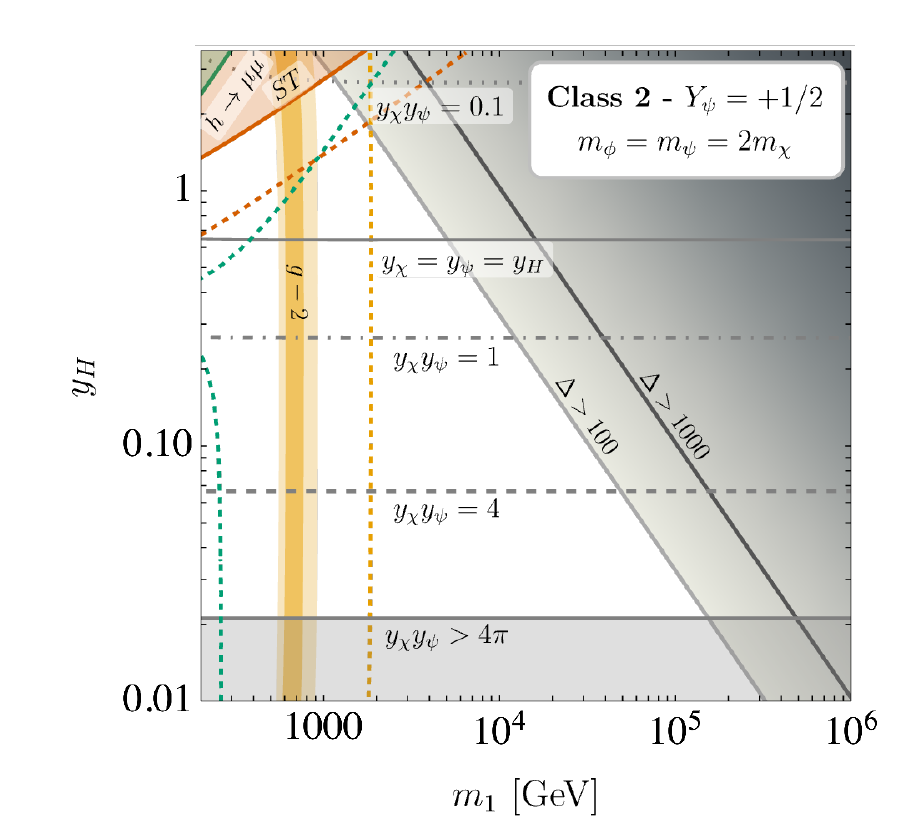}
    \includegraphics[height=0.45\textwidth]{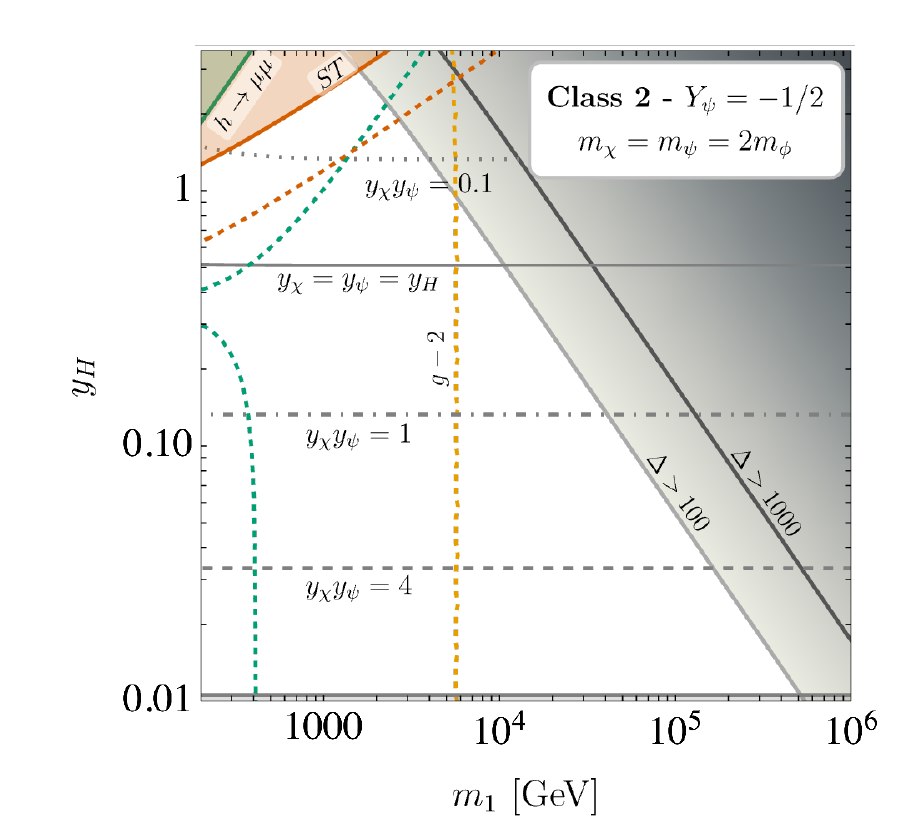}
	\caption{Theoretical (grey) and experimental (coloured) constraints on the Class~2 benchmark model (see~\cref{fig:1mu} for details).  The product of Yukawas required to reproduce the observed muon mass is shown by contours at $y_\chi y_\psi \in \{0.1, 1, 4, 4\pi\}$ and we also show a contour where $y_\chi = y_\psi = y_H$. 
	}
	\label{fig:2mu}
\end{figure}

In \cref{fig:2mu} we show the theoretical and experimental constraints on these models, after fixing the product of Yukawa couplings $y_\chi y_\psi$ to reproduce the muon mass\footnote{Only the sub-leading contributions to $(g-2)_\mu$ depend on the individual Yukawas; here we take $y_\chi=y_\psi$.}.  In the left panel the lightest exotic is $\chi$-like, while in the right panel we set the fermion mass parameters equal and $\phi$ is the lightest exotic.  Perturbative unitarity requires $y_H, y_\chi, y_\psi \lesssim \sqrt{4\pi}$; combining this with the requirement that the observed muon mass is obtained restricts $y_H \gtrsim 0.02$ on the left and $y_H \gtrsim 0.01$ on the right.  The difference is driven by the choice of parameter slice rather than the differing hypercharges between the models.  In addition to contours of constant $y_\chi y_\psi$ we show where $y_H = y_\chi = y_\psi$; for the left slice this occurs at $y_H \approx 0.6$ and for the right at $y_H \approx 0.5$.  We see that fine-tuning of the Higgs mass, $\Delta$, becomes significant for masses above $\approx10^4\,\text{GeV}$ for $y_H$ of $\mathcal{O}(1)$ and $\approx10^5\,\text{GeV}$ for $y_H$ of $\mathcal{O}(0.1)$.

The left plot shows that the $(g-2)_\mu$ anomaly can be reproduced for the $Y_\psi = +1/2$ model for $m_1 \approx 600\,\text{GeV}$, almost independently of $y_H$. For large $y_H$ and a light new physics scale there is a slight dependence on $y_H$, due to the fact that $m_{1,2}$ in \cref{eq:2-g-2} depend on $y_H$.  If the anomaly disappears, future measurements of $(g-2)_\mu$ will probe $m_1 \lesssim 2\,\text{TeV}$.  The contribution to $(g-2)_\mu$ in \cref{eq:2-g-2} can only have the correct sign to explain the anomaly when $\overline{Q}_F > -1/2$. Hence, the model in the right-hand panel is currently in tension with the experimental result.  If the anomaly does not persist, future measurements will probe the model up to $m_1 \approx 6\,\text{TeV}$.

In these models the current constraints from the $S$ and $T$ parameters are stronger than the Higgs-muon signal strength.  In the $Y_\psi = +1/2$ slice they rule out the explanation of the $(g-2)_\mu$ anomaly above $y_H \approx 2$, while future electroweak~\cite{1608.01509} and Higgs~\cite{deBlas:2019rxi} measurements at the FCC could probe down to $y_H \approx 1$.  For both slices, if the anomaly disappears then future $(g-2)_\mu$ measurements at Fermilab will provide a better probe of the parameter space than electroweak or Higgs measurements.  Current and future measurements of deviations in the $Z$-muon couplings~\cite{deBlas:2019wgy} are significantly weaker than the other constraints, so we do not show them.

As was the case for the Class~1 models, there are collider bounds on the exotics from smuon searches~\cite{1908.08215}. For the $Y_\psi=+1/2$ model, one obtains $m_\phi \gtrsim 450\,\text{GeV}$ (for $m_1 \lesssim 150\,\text{GeV}$). This search could be recast to give similar bounds on the mixed fermion states in the $Y=-1/2$ model. These searches assume that $U(1)_X$ is (approximately) unbroken and the lightest exotic is collider-stable. The reach of direct searches at future colliders is also comparable to Class~1 models.

Both of these models contain a dark matter candidate, unless the $U(1)_X$ symmetry is completely broken.  In the $Y_\psi = +1/2$ model the lightest mixed $\psi-\chi$ state may be dark matter.  It can be produced via freeze-out through annihilation into muons (via $t$-channel $\phi$ exchange), electroweak bosons, or co-annihilation processes if the $m_1-\phi$ mass splitting is less than $\sim 30\%$.  Unless the mixing angles $\theta_L$ and $\theta_R$ are very small there will be strong direct detection limits due to $Z$-exchange.  These may, however, be avoided if $\chi$ is Majorana or if a Dirac $\chi$ is split by a Majorana mass term.  In~\cite{Kowalska:2020zve} a similar model, without the chiral symmetries $S_\chi$ and $S_\psi$ and with a tree-level muon mass, is shown to have viable parameter space which fits the $(g-2)_\mu$ anomaly and has a phenomenologically successful dark matter candidate.  It would be very interesting to investigate whether these successes could be combined with radiative muon mass generation.

In the $Y_\psi = -1/2$ model the scalar $\phi$ may be a dark matter candidate.  Its freeze-out abundance would be determined by $t$-channel $\psi$ and $\chi$ exchange and possibly co-annihilation.  While a complex scalar would have stringent constraints from direct detection due to the one-loop photon penguin diagram, these weaken considerably for real scalar dark matter where the leading processes occur at two-loop~\cite{0907.3159,Kawamura:2020qxo}.  A similar model, without the chiral symmetries and with a tree-level muon mass, can fit the $(g-2)_\mu$ anomaly and provide a viable dark matter candidate~\cite{Kowalska:2017iqv,Kawamura:2020qxo,Kowalska:2020zve}.

The phenomenology is expected to be broadly similar in non-minimal models. The Higgs-muon effective Yukawa \eqref{eq:2-eff-yuk} is model-independent, and although there can be new contributions to the Higgs-gluon coupling these are generally small in Class~2 models (see Ref.~\cite{Baker:2020vkh}).  The $(g-2)_\mu$ contribution depends on the model only through $\overline{Q}_F$ and is expected to be of a similar magnitude in all models of this class. The contribution in \cref{eq:2-g-2} has the correct sign to fit the anomaly when $\overline{Q}_F \geq 0$, and can have either sign when $ -1/2 < \overline{Q}_F < 0$.  The most significant difference will be in the bounds from direct searches if the exotics are coloured.

\section{Conclusions}
\label{sec:conclusions}

With the first evidence that the Higgs couples to muons, it is an important time to re-evaluate the alternatives to the SM mechanism of fermion mass generation. Radiative models, in particular, present a logical and well-motivated possibility. In this work, we have explored the minimal models that radiatively generate the muon mass at one-loop. Assuming only new exotic scalars and vector-like fermions, these models fall into two general classes. By considering several benchmark models in detail, we have verified that the muon mass can indeed be radiatively generated with $\mathcal{O}(0.1-1)$ couplings while remaining consistent with all current experimental results.

Radiative mass generation is inevitably associated with new contributions to anomalous magnetic moments. We have demonstrated that these contributions are largely insensitive to the couplings of the exotic fields, with the leading contribution depending only on the masses of the exotics.  For the benchmark models we studied, we found that $(g-2)_\mu$ provides the most sensitive probe of these models, and is likely to do so for the foreseeable future. Furthermore, the current anomaly could be a first hint that the muon mass has a radiative origin.

Radiative models also predict deviations in the Higgs-muon coupling, violating the SM relation $y_f = \sqrt{2} m_f/v$. However, we have found that the deviation is small across much of the parameter space and the reach of Higgs measurements is limited. This is the case even with the precision of future colliders such as FCC. Direct searches for the exotics are more promising and, while the current bounds from the LHC are relatively weak, future hadron and lepton colliders could be competitive with $(g-2)_\mu$ measurements and will be particularly interesting if the current anomaly persists.

The minimal models all possess a $U(1)_X$ symmetry that stabilises the lightest exotic. This suggests a natural connection between radiative mass generation and dark matter. Each of the four benchmark models we studied features a potential dark matter candidate. Previous studies have already confirmed that these (or closely related) models can yield the observed relic abundance while satisfying other bounds, although not necessarily in the regions of parameter space relevant for radiative mass generation. It would be interesting to revisit the dark matter phenomenology of these models with radiative mass generation in mind.

\section{Acknowledgements}

The authors would like to thank John Gargalionis for useful discussion and collaboration in the initial stages of this project.  This work was supported by the Australian Government through the Australian Research Council.

\bibliography{paper}
\bibliographystyle{JHEP}

\end{document}